\documentclass[10pt,a4paper,final]{iopart}
\usepackage{bm}
\usepackage{iopams}
\usepackage{graphicx}
\usepackage[breaklinks=true,colorlinks=true,linkcolor=blue,urlcolor=blue,citecolor=blue]{hyperref}
\usepackage{subcaption}
\usepackage{bbm}
\usepackage[dvipsnames]{xcolor}
\expandafter\let\csname equation*\endcsname\relax
\expandafter\let\csname endequation*\endcsname\relax
\usepackage{amsmath,amsfonts,amssymb,amsthm}
\usepackage{braket}
\newtheorem{theorem}{Theorem}[section]
\def\Tr{\hbox{Tr}}

\newcommand{{\lambdaB}}{\bm\lambda}
\newcommand{\QFIbm}{\bm{\mathcal{Q}}}
\newcommand{\UCbm}{\bm{\mathcal{U}}}

\begin{document}

\title[Asymptotic incompatibility in multiparameter quantum estimation]{On the properties of the asymptotic incompatibility measure in multiparameter quantum estimation}

\author{Alessandro Candeloro$^{1,2}$, Matteo G.A. Paris$^{1,2}$, Marco G. Genoni$^{1,2}$}
\address{$^1$ Quantum Technology Lab, Dipartimento di Fisica 
Aldo Pontremoli, Universit\`a degli Studi di Milano, I - 20133, 
Milano, Italy \\ 
$^2$ Istituto Nazionale di Fisica Nucleare, Sezione di Milano, I - 20133, 
Milano, Italy}
\eads{\mailto{alessandro.candeloro@unimi.it}, \mailto{matteo.paris@fisica.unimi.it},\mailto{marco.genoni@fisica.unimi.it}}
\begin{abstract}
We address the use of {\em asymptotic incompatibility} (AI) to assess the {\em quantumness} of a multiparameter quantum statistical model. AI is a recently introduced measure which quantifies the difference between the Holevo and the 
SLD scalar bounds, and can be evaluated using only the symmetric logarithmic derivative (SLD) operators of the model. At first, we evaluate analytically the AI of 
the most general quantum statistical models involving two-level (qubit) and single-mode Gaussian continuous-variable quantum systems, and prove that AI is a simple monotonous function of the state purity. Then, we numerically investigate the same problem for qudits ($d$-dimensional quantum systems, with $2 < d \leq 4$), showing that, while in general AI is not in general a function of purity, we have enough numerical evidence to conclude that the maximum amount of AI is attainable only for quantum statistical models characterized by a purity larger than $\mu_{\sf min} = 1/(d-1)$. In addition, by parametrizing qudit states as thermal (Gibbs) states, numerical results suggest that, once the spectrum of the Hamiltonian is fixed, the AI measure is in one-to-one correspondence with the fictitious temperature parameter $\beta$ characterizing the family of density operators. Finally, by studying in detail the definition and properties of the AI measure we find that: i) given a quantum statistical model, one can readily identify the maximum number of {\em asymptotically compatibile} parameters; ii) the AI of a quantum statistical model bounds from above the AI of any sub-model that can be defined by fixing one or more of the original unknown parameters (or functions thereof), leading to possibly useful bounds on the AI of models involving noisy quantum dynamics.
\end{abstract}

\vspace{2pc}

\submitto{\jpa}
\section{Introduction}
Quantum metrology is currenty revolutionizing the field of parameter estimation and sensing by enhancing precision to a level that cannot be achieved via purely classical means \cite{Giovannetti2011,Demkowicz-Dobrzanski2015a,Degen2016,Pirandola2018}. Most of the paradigmatic results in quantum metrology have been obtained for the estimation of single parameters, typically corresponding to a phase or a frequency. In those cases, the ultimate and in principle achievable bounds are derived in terms of the quantum Fisher information, which is in turn obtained via the symmetric logarithmic derivative (SLD) operator \cite{helstrom1976quantum,Paris2009}. The extension to the multiparameter case is however not straightforward \cite{albarelli2020perspective,Szczykulska2016,liu2019quantum,sidhu2020geometric}. In fact, besides the expected complications due to the fact that one needs to estimate more than one parameter, the peculiar properties of quantum mechanics make this extension definitely non-trivial. In particular, the (potential) non-commutativity of the optimal measurements corresponding to different parameters poses more fundamental and interesting questions. A bound to precision of multiparameter estimation may be obtained by generalizing the single-parameter case, i.e. using the SLD operators corresponding to the different parameters. On the other hand, several other bounds may be introduced, the most relevant being the ones based on the right logarithmic derivative operator \cite{Yuen1973}, on an operator interpolating between right and symmetric logarithmic derivative \cite{Yamagata2021}, and the so-called Holevo bound \cite{holevo2011probabilistic}. 

In general, all these bounds are not tight, i.e. it is not guaranteed that there 
exists a quantum measurement that allows to saturate them. However, the Holevo bound stands out as the most fundamental (scalar) bound for quantum multiparameter estimation as it may be saturated by performing collective measurements on an asymptotically large number of copies of the quantum state encoding the set of parameters~\cite{Hayashi2008a,Guta2006,Yang2019} (tighter bounds that apply in the scenario where one allows measurements on a finite number of copies have been derived more recently \cite{Conlon2021,chen2021}). Moreover, in some cases, e.g. for pure states \cite{Matsumoto2002} and for the estimation of a displacement by Gaussian probes \cite{holevo2011probabilistic}, the Holevo bound can be achieved also in the standard scenario, i.e. by single-copy measurements.
These fundamental aspects have been investigated in several multiparameter problems that may have practical applications in the quantum regimes, such as superresolution of incoherent sources \cite{Chrostowski2017,Yu2018,Napoli2018,Fiderer2021}, estimation of multiple phases (or in general of unitary parameters) \cite{Ballester2004,Vaneph2012,Genoni2013b,Humphreys2013,Baumgratz2015,Gagatsos2016a,Knott2016,Pezze2017,Roccia2017,bradshaw2017tight,bradshaw2018ultimate,albarelli2019evaluating,friel2020}, and estimation of phase and noise \cite{Vidrighin2014,Crowley2014,Altorio2015a,Roccia2017,Roccia2018,albarelli2019evaluating}. 

Recently, the difference between the SLD and the Holevo bound led to more fundamental studies, focusing on the incompatibility of parameters and leading to the definition of measures able to capture this particular aspect \cite{carollo2019quantumness, Belliardo2021}.
In particular in \cite{carollo2019quantumness}, a quantity has been introduced to quantify the {\em quantumness} of a quantum multiparameter problem. This quantity, denoted by $\mathcal{R}_{\bm{\lambda}}$,  is equal to zero if and only if the Holevo bound coincides with the SLD based bound, and thus if the parameters to be estimated are {\em asymptotically compatible}. Similarly, it takes its largest value $\mathcal{R}_{\bm{\lambda}}=1$ when the difference between the Holevo and the SLD bound is maximum, that is when the parameters are highly incompatible even in the asymptotic regime. 
Given these properties, in the following we will refer to $\mathcal{R}_{\bm{\lambda}}$ 
as {\em asymptotic incompatibility} (AI). The properties of AI has been discussed in relationship with the phase diagram of a quantum many-body system in \cite{carollo2019quantumness}, while its main properties and its behaviour for estimation problems encoded in qubit systems have been thoroughly investigated in \cite{razavian2020quantumness}.

In this paper, we focus on the fundamental properties of the AI measure $\mathcal{R}_{\bm{\lambda}}$ in local quantum estimation theory. We study its relationship with the purity of the density operators defining the quantum statistical model, and prove that for some classes of states, the two quantities are indeed in one-to-one correspondence. More in general, we find evidence that the maximum amount of incompatibility may be achieved only for purity exceeding a threshold depending on the dimension of the Hilbert space. We conjecture that this may be a general property of quantum statistial  models. Moreover, we derive some bounds on the AI measure for quantum statistical {\em submodels}, and provide a method to directly identify the maximum number of compatible parameters in certain estimation problems.
 
The manuscript is organized as follows: in Sec. \ref{s:QET} we review the main aspects of multiparameter quantum estimation and we introduce the AI measure $\mathcal{R}_{\bm{\lambda}}$. In Sec. \ref{s:AIandPurity}, we show our main results concerning the relationship between the AI measure and the purity of the quantum states, while in Sec. \ref{s:properties} we will derive further properties regarding the AI of parameters of specific quantum statistical models. Sec. \ref{s:conclusion} closes the paper  with some concluding remarks.
\section{Local quantum multiparameter estimation}\label{s:QET}
In this section we briefly review the tools of local quantum multiparameter estimation. The goal of a typical metrological problem is to estimate the value of a set $\bm\lambda$ of $\mathfrak{p}$ independent parameters from a set of $M$  outcomes $\bm{x} = \{x_1,x_2,..., x_M\}$ that are distributed according to the conditional probability distribution $p(x\vert\bm\lambda)$, usualy referred to as statistical model. The estimated value of the parameters is obtained via an estimator function $\bm{\hat{\lambda}}(\bm{x})$. If we denote by $\mathbb{E}_{\bm\lambda}[\bm{\hat{\lambda}}]$ the mean value of the function $\bm{\hat{\lambda}}$ over the probability distribution $p(x|\bm\lambda)$, then the precision of any unbiased estimator is quantified by the covariance matrix $\textup{\textbf{Cov}}_{\bm{\lambda}} [\bm{\hat{\lambda}}]_{\alpha\beta} = \mathbb{E}_{\bm{\lambda}} \left[(\bm{\hat{\lambda}}-\mathbb{E}_{\bm{\lambda}}[\bm{\hat{\lambda}}])_{\alpha}(\bm{\hat{\lambda}}-\mathbb{E}_{\bm{\lambda}}[\bm{\hat{\lambda}}])_{\beta}\right]$. The latter satisfies the classical matrix Cram\'er-Rao (CCRB) bound 
\begin{eqnarray}
 	\textup{\textbf{Cov}}_{\bm{\lambda}}[\hat{\bm{\lambda}}] \geq \frac{1}{M}\bm{\mathcal{F}}^c(\bm{\lambda})^{-1},
 	\label{CFIM:eq}
\end{eqnarray} 
where the classical Fisher Information matrix (CFIM) is defined via its elements
\begin{eqnarray}
	\mathcal{F}^c_{\alpha \beta}(\bm{\lambda}) = \int_{\chi} {\rm d} x \,   p(x \vert \bm{\lambda})\, \left[ \partial_\alpha \log{p(x \vert \bm{\lambda})} \right]\, \left[\partial_\beta  \log{p(x \vert \bm{\lambda})}\right]\,.
\end{eqnarray}
The inequality must be understood as the fact that $\textup{\textbf{Cov}}_{\bm{\lambda}}[\hat{\bm{\lambda}}] - M^{-1}\bm{\mathcal{F}}^c(\bm{\lambda})^{-1}$ is a semi-definite positive matrix \cite{amari2007methods,cramir1946mathematical,hayashi2016quantum}. Classical estimation theory studies the conditions for the attainability of the CCRB and optimal estimators are the ones for which the inequality is saturated. In the limit of a large number of measurements $M\to\infty$, the maximum likelihood estimator is proven to be optimal \cite{lehmann2006theory}.
\par
Moving to the quantum realm, the parameters are encoded in a family of density operators $\varrho_{\bm{\lambda}}$ referred to as a quantum statistical model. We remark that in the following we will restrict ourselves to quantum statistical models described by quantum states $\varrho_{\bm{\lambda}}$ whose rank in the Hilbert space does not change by varying the parameters $\bm\lambda$ in the allowed region of the parameters' space. This will allow us to avoid discontinuities in the behaviour of the figures of merit we will consider \cite{Safranek2017,Seveso2019}. The conditional probabilities of observing a certain outcome are obtained through the Born rule $p(k\vert \bm{\lambda}) = \Tr[\varrho_{\bm{\lambda}}\hat{\Pi}_k]$. These depends on the measurement process involved, which is described in its generality by a POVM, i.e by a set of operators  $\boldsymbol{\Pi} = \{\hat{\Pi}_k \vert \hat{\Pi}_k\geq 0, \sum_k \hat{\Pi}_k = \hat{\mathbb{I}}\}$.  By using different approaches, measurement-independent bounds on the matrix covariance have been introduced \cite{albarelli2020perspective,liu2019quantum,sidhu2020geometric}. Among them, in this paper we are going to deal first with the bound obtained via the symmetric logarithmic derivative (SLD) operators, that is typically referred to as SLD-bound. SLD operators $\hat{L}^S_\alpha$, are defined implicitly as the solutions of the Lyapunov equations 
\begin{eqnarray}
	\partial_\alpha \varrho_{\bm{\lambda}} = \frac{\hat{L}^S_\alpha\varrho_{\bm{\lambda}} + \varrho_{\bm{\lambda}} \hat{L}^S_\alpha}{2}. \end{eqnarray}
The corresponding SLD quantum Fisher information matrix (QFIM) is obtained via the formula 
\begin{eqnarray}
	\label{SLD-QFIM:eq}
	\mathcal{Q}_{\alpha\beta}(\bm{\lambda}) = \Tr\left[\varrho_{\bm{\lambda}} \frac{\hat{L}_\alpha^S\hat{L}_\beta^S+\hat{L}_\beta^S\hat{L}_\alpha^S}{2}\right],
\end{eqnarray}
and the corresponding SLD quantum Cramer-Rao (QCR) matrix bound reads \cite{hayashi2016quantum}, 
\begin{eqnarray}
	\label{SLD-QCRB:eq}
	\textup{\textbf{Cov}}_{\bm{\lambda}}[\hat{\bm{\lambda}}] \geq \bm{\mathcal{Q}}(\bm{\lambda})^{-1}. 
\end{eqnarray}
It can be shown that his bound is in fact achievable for single-parameter estimation \cite{helstrom1976quantum,braunstein1994statistical,paris2009quantum}. This means that for a single-copy of the state $\varrho_{\bm\lambda}$, a POVM whose Fisher Information is equal to the SLD-QFI exists, and remarkably one can prove that this optimal POVM corresponds to the projectors over the eigenstates of the SLD operator. If we now consider the multiparameter case $\mathfrak{p}>1$, the different SLD operators may not commute, and an optimal simultaneous estimation of all the parameters can not be in general obtained. As a consequence, the QCR matrix bound may not always be achievable in the single-copy scenario and more informative bounds than the one provided by the SLD operators may exist. 
\par 
When dealing with matrix inequalities, an additional problem arises already in classical estimation theory, i.e. the order of the CFIM is partial. This means that, given two experimental strategies specified by two distinct POVM $\bold{\Pi}^{(1)}$ and $\bold{\Pi}^{(2)}$ to which corresponds $\bm{\mathcal{F}}^{(1)}$ and $\bm{\mathcal{F}}^{(2)}$, it might be that both $\bm{\mathcal{F}}^{(1)}\ngeq \bm{\mathcal{F}}^{(2)}$ and $\bm{\mathcal{F}}^{(2)}\ngeq \bm{\mathcal{F}}^{(1)}$. Thus, to understand which strategies is better, scalar bounds have been introduced. Here, we study the one defined in terms of a weight matrix $\bm{\mathcal{W}}$, a real and positive definite matrix of dimension $\mathfrak{p}\times \mathfrak{p}$. The SLD-QCR scalar bound obtainable from \Eref{SLD-QCRB:eq} can be written as follow
\begin{eqnarray}
	\label{scalar-SLD-QCRB:eq}
	\Tr\left[\bm{\mathcal{W}}\,\textup{\textbf{Cov}}_{\bm{\lambda}}[\hat{\bm{\lambda}}]\right] \geq C^S(\bm{\lambda},\bm{\mathcal{W}})  = \Tr\left[\bm{\mathcal{W}} \bm{\mathcal{Q}}(\bm{\lambda})^{-1}\right].
\end{eqnarray} 
The role of the weight matrix $\bm{\mathcal{W}}$ is to balance the precision of different parameters. In addition, there is a one to one correspondence between $\bm{\mathcal{W}}$ and a change of parametrization. Given a new set of parameter $\bm\gamma = \bm\gamma(\bm\lambda)$ defined in terms of the previous one, the changes in the QFI-SLD matrix is determined by the reparametrization matrix $\mathcal{B}_{\alpha\beta}= \partial\lambda_\beta/\partial\gamma_\alpha$. The SLD-QFI matrix for the new parameter is \cite{gge08}
\begin{eqnarray}
	\bm{\mathcal{Q}}(\bm\gamma) = \bm{\mathcal{B}} \bm{\mathcal{Q}}(\bm\lambda)  \bm{\mathcal{B}}^T.
\end{eqnarray}
Besides, any real positive definite matrix can be decomposed as $\bm{\mathcal{W}} = \bm{\mathcal{L}}\bm{\mathcal{L}}^{T}$, where $\bm{\mathcal{L}}$ is unique. We see that if $\bm{\mathcal{L}}=\bm{\mathcal{B}}^{-1}$, then a particular choice of $\bm{\mathcal{W}}$ corresponds to a unique change of parametrization induced by $\bm{\mathcal{B}}$.
\par
As its matrix counterpart, the SLD-QCR scalar bound \eref{scalar-SLD-QCRB:eq} is not in general attainable due to the incompatibility of the optimal measurements corresponding to the different parameters. The problem of finding the most informative scalar bound was addressed by Holevo in \cite{holevo2011probabilistic}. The solution takes the name of Holevo Cram\'er-Rao bound (HCRB) and it will be denoted by $C^H(\bm{\lambda},\bm{\mathcal{\bm{W}}})$. It represents the most informative bound for the asymptotic model in which a collective measurement is performed on an asymptotically large number of copies of the state $\varrho_{\bm\lambda}^{\otimes n} = \bigotimes_{j=1}^n\varrho_{\bm\lambda}$, with $n\to\infty$ \cite{Hayashi2008a,Guta2006}. In this limit, the bound is indeed achievable and for this reason it is typically considered as the most fundamental scalar bound for quantum multiparameter estimation. However, the evaluation of the HCRB requires a non-trivial minimization (see \ref{app:HCRB} for the formal definition of the HCRB). Nonetheless, some results have been obtained both numerically and even analytically under some assumptions on the quantum statistical model \cite{Crowley2014,Baumgratz2015,suzuki2016explicit,bradshaw2017tight,bradshaw2018ultimate,albarelli2019evaluating,sidhu2021tight,suzuki2019information,Gorecki2019,Assad2020}

\subsection{Asymptotic incompatibility measure}
As observed before, in a multiparameter scenario the optimal strategies for each single parameter estimation may not be compatible and as a result the lower bound \eref{scalar-SLD-QCRB:eq} can not always be attained. A key object in this respect is the commutator $[ \hat{L}_\alpha^S , \hat{L}_\beta^S]$ between the different SLD-operators: if this commutator is equal to zero for all the parameters in $\boldsymbol{\lambda}$, 
all the SLD-bounds for each single parameter are simultaneously achievable in the single-copy scenario by performing the same POVM, and as a result, both the matrix and scalar SLD-QCR bound given in Eq. \eqref{SLD-QCRB:eq} and \eqref{scalar-SLD-QCRB:eq}. These models are known in the litterature as quasi-classical model \cite{albarelli2020perspective,suzuki2019information}. However, as we will describe in a moment, its average value over the quantum statistical model $\varrho_{\boldsymbol{\lambda}}$ plays an important role too. Typically this quantity is introduced via the so-called Uhlmann (or Berry) curvature \cite{carollo2018uhlmann,liu2019quantum}, defined via its matrix elements
\begin{eqnarray}
	\label{Uhlmann-matrix:eq}
	\mathcal{U}_{\alpha\beta}(\bm{\lambda}) := - \frac{i}{2} \Tr\left[\varrho_{\bm{\lambda}}\left[\hat{L}^S_\alpha,\hat{L}^S_\beta\right]\right].
\end{eqnarray}
One proves that a necessary and sufficient condition for the attainability of the SLD-QCR scalar bound \eref{scalar-SLD-QCRB:eq} in the asymptotic limit is given by the compatibility condition, or weak commutativity, $\bm{\mathcal{U}} = \bm{0}$ \cite{ragy2016compatibility}. If this condition is fulfilled, that is if all the SLD operators commute on average, then $C^H(\bm\lambda,\bm{\mathcal{W}}) = C^S(\bm\lambda,\bm{\mathcal{W}})$, i.e. the SLD-QCR scalar bound can be attained in the asymptotic model $\varrho_{\boldsymbol{\lambda}}^{\otimes n}$ with $n\to \infty$. In the following, we will refer to models that satisfy this condition as asymptotically classical model and parameters belonging to these models as asymptotic compatible parameters \cite{albarelli2020perspective,suzuki2019information}.
\par
More recent results have shown how the HCRB can be bounded from above as 
\begin{eqnarray}
 	C^S(\bm{\lambda},\bm{\mathcal{W}})  \leq C^H(\bm{\lambda},\bm{\mathcal{W}}) \leq (1 + \mathcal{R}_{\bm\lambda}) C^S(\bm{\lambda},\bm{\mathcal{W}}) \leq 2 \, C^S(\bm{\lambda},\bm{\mathcal{W}})\,
 	\label{eq:rangeHolevoBound}
 \end{eqnarray} 
that is $C^H(\bm{\lambda},\bm{\mathcal{W}})$ cannot be larger than two times the SLD QCR scalar bound. In the chain of inequalities above, we have introduced the following parameter
\begin{align}
\mathcal{R}_{\bm\lambda} = \lvert\vert i \,\bm{\mathcal{Q}}(\bm{\lambda})^{-1} \bm{\mathcal{U}}(\boldsymbol{\lambda}) \rvert\rvert_\infty \, , \label{eq:Rdefinition}
\end{align}
where $\lvert\lvert {\bf A} \rvert\rvert_\infty$ denotes the largest eigenvalue of the matrix ${\bf A}$. In the rest of the manuscript we will focus on this quantity, that has been introduced, as a {\em quantumness} parameter, in \cite{carollo2019quantumness} and studied in detail for qubit systems in \cite{razavian2020quantumness}. As already suggested in Eq. (\ref{eq:rangeHolevoBound}), one can prove that 
$$
0 \leq \Delta C(\lambdaB,\mathbf W) \leq \mathcal{R}_{\bm\lambda} \leq 1\,,
$$
where we have defined the renormalized difference between the Holevo and SLD-bound
\begin{align}
\Delta C(\lambdaB ,{\bf W}) = \frac{C^{\sf H}(\lambdaB , {\bf W} )-C^{\sf S}(\lambdaB , {\bf W} )}{C^{\sf S}(\lambdaB , {\bf W} )} \,.
\end{align}
Moreover one also proves two relevant properties \cite{carollo2019quantumness,razavian2020quantumness}: (i) $\mathcal{R}_{\bm\lambda} = 0$ if and only if $\bm{\mathcal{U}}(\boldsymbol{\lambda}) = 0$, that is when the model is asymptotically classical; (ii) $\mathcal{R}_{\bm\lambda}$ is a property of the quantum statistical model $\varrho_{\boldsymbol{\lambda}}$ only, being also independent on possible reparametrization and, as a consequence, on the weight matrix $\bm{\mathcal{W}}$. In \cite{razavian2020quantumness} it was also studied its relationship with the quantity 
\begin{align}
\Delta C_{\sf max} = \max_{{\mathbf W} >0} \Delta C({\lambdaB},{\mathbf W}) \,,
\end{align}
for quantum statistical models encoded in qubit systems. The quantity $\Delta C_{\sf max}$ stands out as a natural measure of {\em asymptotic incompatibility} (AI) and it is in general upper bounded by $\mathcal{R}_{\bm\lambda}$. It was shown that, while for several quantum statistical model the bound is indeed tight, that is one finds $\Delta C_{\sf max}=\mathcal{R}_{\bm\lambda}$, there exist  counterexamples where $\Delta C_{\sf max}$ is strictly smaller than $\mathcal{R}_{\bm\lambda}$. However, also in these examples one observes that the two quantities have the same general behaviour and in general that the order relations induced by them are equivalent. \\
For all these reasons, and also considering the fact that the evaluation of the quantity $\mathcal{R}_{\bm\lambda}$ is quite straightforward and, at difference with $\Delta C_{\sf max}$, does not rely on the evaluation of the Holevo bound and on its (non trivial) maximization over the weight matrices $\mathbf W$, in the following  we will restrict to it and we will in general refer to $\mathcal{R}_{\bm\lambda}$ as a {\em AI measure} of the quantum statistical model $\varrho_{\bm\lambda}$. %
\section{Asymptotic incompatibility and purity of the quantum statistical model}\label{s:AIandPurity}
In this section we discuss the relationship between the AI measure $\mathcal{R}_{\lambdaB}$ and the purity of the quantum state $\varrho_{\bm\lambda}$ corresponding to the quantum statistical model under exam. We will mainly focus to scenarios where $\varrho_{\bm\lambda}$ describes the most general quantum state of a particular quantum system, starting from the simplest cases of a qubit and of a single-mode Gaussian continuous-variable quantum state, and then moving our attention to qudits, that is to quantum states living in a $d$-dimensional Hilbert space.
\subsection{Asymptotic incompatibility of state parameters in full tomography of qubit systems}
A generic mixed qubit state is typically written as 
\begin{equation}
\varrho_{\bm\lambda} = \frac{1}{2}\left(\mathbbm{1} + \sum_{j=1}^3 \gamma_j \sigma_j \right) \,,
\end{equation}
where the matrices $\sigma_j$ denote the Pauli matrices, while $\gamma_1 = r \sin\theta \cos\phi$, $\gamma_2 = r \sin\theta \sin\phi$, $\gamma_3 = r \cos\theta$. By considering the set of $\mathfrak{p}=3$ parameters $\lambdaB = ( r ,\theta, \phi)$ characterizing the vector in the Bloch sphere, one can readily derive the SLD operators by solving the corresponding Lyapunov equations, and obtains the SLD-QFI and the Ulhman curvature matrices
 \begin{align}
\QFIbm(\lambdaB)&=\left(
\begin{array}{ccc}
1/(1-r^2)&0 & 0\\
0& r^2 & 0 \\
0 & 0 & r^2 \sin^{2}\theta 
\end{array}
\right)\,,\\ 
\UCbm(\lambdaB) &= \left(
\begin{array}{ccc}
0 & 0 & 0 \\
0 & 0 &  r^3 \sin\theta \\
0 & - r^3 \sin\theta&0 
\end{array}
\right)\,.
\end{align}
 The corresponding AI measure has been already derived in \cite{razavian2020quantumness}, obtaining $\mathcal{R}_{\rm qubit} = r$. The purity of the quantum state $\varrho_{\lambdaB}$ can be easily evaluated, obtaining $\mu=\Tr[\varrho_{\lambdaB}^2]=(1+r^2)/2$, and thus leading to the result
 \begin{align}
\label{eq:AIqubit}
 \mathcal{R}_{\rm qubit}=  \sqrt{2\mu -1} \,.
 \end{align} 
We thus observe that the AI for the full tomography $\mathcal{R}_{\rm qubit} $ is indeed a monotonous function of the purity of the quantum state, and that in particular it takes its limiting values $\mathcal{R}_{\rm qubit}=0$ for the maximally mixed state, and $\mathcal{R}_{\rm qubit}=1$ for pure states only.\\
\subsection{Asymptotic incompatibility for a single-mode continuous-variable Gaussian system}
As a second example we now consider a single-mode continuous-variable quantum system, described by annihilation and creation operators satisfying the canonical commutation relation $[\hat{a},\hat{a}^\dagger ] = \mathbbm{1}$. Quantum states describing such systems belongs to an infinite-dimensional Hilbert space. A well known subclass of such states is given by Gaussian states, typically defined as those states having a Gaussian Wigner function \cite{ferraro2005gaussian,serafini2017quantum}. These states have been indeed studied in great detail, both for the simple mathematical formulation (they can be fully described by first and second moments of the quadrature operators $\hat{q} = (\hat{a} + \hat{a}^\dag)/\sqrt{2}$ and $\hat{p} = i(\hat{a}^\dagger-\hat{a} )/\sqrt{2}$) and for practical and fundamental reasons (they can be easily generated in the lab, and they can be exploited for implementing quantum technology protocols as quantum teleportation). We leave more details on Gaussian states in \ref{a:gaussian}, along with the formulas needed to evaluate the SLD-QFI and the Uhlmann curvature matrices.\\
We here address the problem of complete estimating an arbitrary single mode Gaussian state, which can be parametrized by $\mathfrak{p}=5$ parameters, resulting from the application of a complex squeezing operator $S(\xi) = \exp\left(\frac{1}{2}\xi (a^\dag)^2 - \frac{1}{2}\xi^*a^2\right)$ and a complex displacement operator $D(\alpha) = \exp\left(\alpha a^\dag-\alpha^*a\right)$ on a thermal state $\nu_N = e^{-\beta \hat{a}^\dagger \hat{a}}/Z$ with average photon number $\Tr[\nu_N \hat{a}^\dag \hat{a}]= N$. The corresponding quantum statistical model is thus represented by the family of density operators \cite{serafini2017quantum}
\begin{eqnarray}
	\varrho_{\lambdaB} = D(\alpha)S(\xi)\nu_N S(\xi)^\dag D(\alpha)^\dag \,, \label{eq:GState}
\end{eqnarray}
with a parametrization in terms of the set $\bm{\lambda} = \{\mathfrak{Re}{\alpha},\mathfrak{Im}{\alpha},r,\varphi,N\}$, and where we have written the two complex parameters as $\xi = r e^{i\varphi}$ and $\alpha = \mathfrak{Re}{\alpha} + i \mathfrak{Im}{\alpha}$. The first moments vector ${\bm d} = (\Tr[\varrho_{\lambdaB}\hat{q}],\Tr[\varrho_{\lambdaB}\hat{p}] )^T$ has elements
\numparts
\begin{eqnarray}
	\Tr[\varrho_{\lambdaB}\hat{q}] =  \sqrt{2}\mathfrak{Re}{\alpha}\,,\\
	\Tr[\varrho_{\lambdaB}\hat{p}] = \sqrt{2}\mathfrak{Im}{\alpha}\,,
\end{eqnarray}
\endnumparts
while the elements of the covariance matrix (CM) (see \ref{a:gaussian} for details) can be expanded in terms of symmetric $2\times 2$ matrices, i.e. $\bm\sigma = s_1\sigma_x + s_2 \bm{I}_2 + s_3\sigma_z$ (with $\sigma_x$ and $\sigma_z$ denoting the standard Pauli matrices), obtaining 
\numparts
\begin{eqnarray}
	s_1 = - (2N+1)\sinh(2r)\sin(\varphi), \\
	s_2 = (2N+1) \cosh(2r), \\
	s_3 = (2N+1) \sinh(2r) \cos(\varphi).	
\end{eqnarray}
\endnumparts
{The matrix elements for the SLD-QFI matrix and for the Uhlmann curvature matrix can be directly evaluated via the formulas 
\begin{align}
 	\mathcal{Q}_{\alpha\beta} &= \frac{1}{2} \frac{\textup{Tr}[(\bm{\sigma}^{-1}\partial_\alpha \bm{\sigma}) (\bm{\sigma}^{-1} \partial_\beta \bm{\sigma})]}{1+\mu^2} + \frac{2\partial_\alpha \mu \partial_\beta \mu}{1-\mu^4} \nonumber \\
& \,\,\,\,\,\,\,\,   + 2\left(\partial_\alpha \bm{d} \right)^T \bm\sigma^{-1} (\partial_\beta \bm{d}) \label{eq:QFIGauss} \\
  \mathcal{U}_{\alpha \beta} & = \frac{\mu^2}{2(\mu^2+1)^2} \textup{Tr}\left[\bm\sigma\Omega \left[\partial_\alpha \bm{\sigma}\bm{\sigma}^{-1}, \partial_\beta \bm\sigma\bm{\sigma}^{-1}\right]\right] \nonumber \\
&\,\,\,\,\,\,\,\,   + 2 \mu^2(\partial_\alpha \bm{d})  \Omega (\partial_\beta \bm{d}), \label{eq:UCGauss}
\end{align}
where the purity of single-mode Gaussian states is given by $\mu = 1/\sqrt{{\rm det}\bm\sigma}=1/(2 N+1)$. Notice that our result differ from that in \cite{nichols2018multiparameter} by a factor $2$ in the first term of the Uhlmann matrix (for a detailed derivation, see \ref{a:gaussian}).}
\par
In particular, for the quantum statistical model defined above, we obtain
\numparts
\begin{eqnarray}
	\label{eq:Qmatelements1}
	\mathcal{Q}_{\mathfrak{Re}{\alpha},\mathfrak{Re}{\alpha}} = 4 \mu \left(\cosh(2r)-\cos(\varphi)\sinh(2r)\right), \\
	\label{eq:Qmatelements2}
	\mathcal{Q}_{\mathfrak{Im}{\alpha},\mathfrak{Im}{\alpha}} = 4 \mu \left(\cosh(2r)+\cos(\varphi)\sinh(2r)\right), \\
	\label{eq:Qmatelements3}
	\mathcal{Q}_{\mathfrak{Re}{\alpha},\mathfrak{Im}{\alpha}} = 4\mu  \sinh(2r)\sin(\varphi), \\
	\mathcal{Q}_{r,r} = \frac{4}{1+\mu^2}, \\
	\label{eq:Qmatelements4}
	\mathcal{Q}_{\varphi,\varphi} = \frac{\sinh(2r)^2}{1+\mu^2},\\
	\label{eq:Qmatelements5}
	\mathcal{Q}_{N,N} =\frac{4\mu^2}{1-\mu^2}, 
	\label{eq:Qmatelements6}
\end{eqnarray}
\endnumparts
and
\numparts
\begin{eqnarray}
	\label{eq:Umatelements1}
	\mathcal{U}_{\mathfrak{Re}{\alpha},\mathfrak{Im}{\alpha}} = 4 \mu^2 \,, \\
	\label{eq:Umatelements2}
	\mathcal{U}_{\varphi,r} = \frac{4\mu \sinh(2r)}{(1+\mu^2)^2},
\end{eqnarray}
\endnumparts
while the matrix elements not reported are null. We emphasize that these values do not depend on $(\Re{\alpha},\Im{\alpha})$, and we also osberve that the SLD-QFI matrix is block diagonal. In particular, the second block regarding the parameters $\left(r,\varphi,N\right)$ is exactly diagonal. Moreover, since most of the entries of the Uhlmann matrix are equal to $0$, we conclude that most of the parameters are orthogonal globally with respect to the SLD-QFI, with the exception of the pairs $(\Re{\alpha},\Im{\alpha})$ and $(r,\varphi)$.
\par
The AI measure can be readily evaluated, obtaning
\begin{align}
\mathcal{R}_{\sf GS} = \frac{2\mu}{1 + \mu^2} \,. \label{eq:AIGauss}
\end{align}
Remarkably, we see that the contribution depending on $(r,\varphi)$ cancel out in the AI $\mathcal{R}_{\sf GS}$. Also in this case we have thus obtained that AI is a monotonous function of purity $\mu$, and that one obtains that all the parameters become asymptotically compatible only in the limit of a maximally mixed state (that is $\mathcal{R}_{\sf GS} =0$ only for $\mu \rightarrow 0$), while the maximum value of incompatibility $\mathcal{R}_{\sf GS}  =1$ is obtained if and only if $\varrho_{\lambdaB}$ is a pure state. 
\subsection{Asymptotic incompatibility for of state parameters in full tomography in qudit systems}
\par
\begin{figure}
\centering
\includegraphics[width=\textwidth]{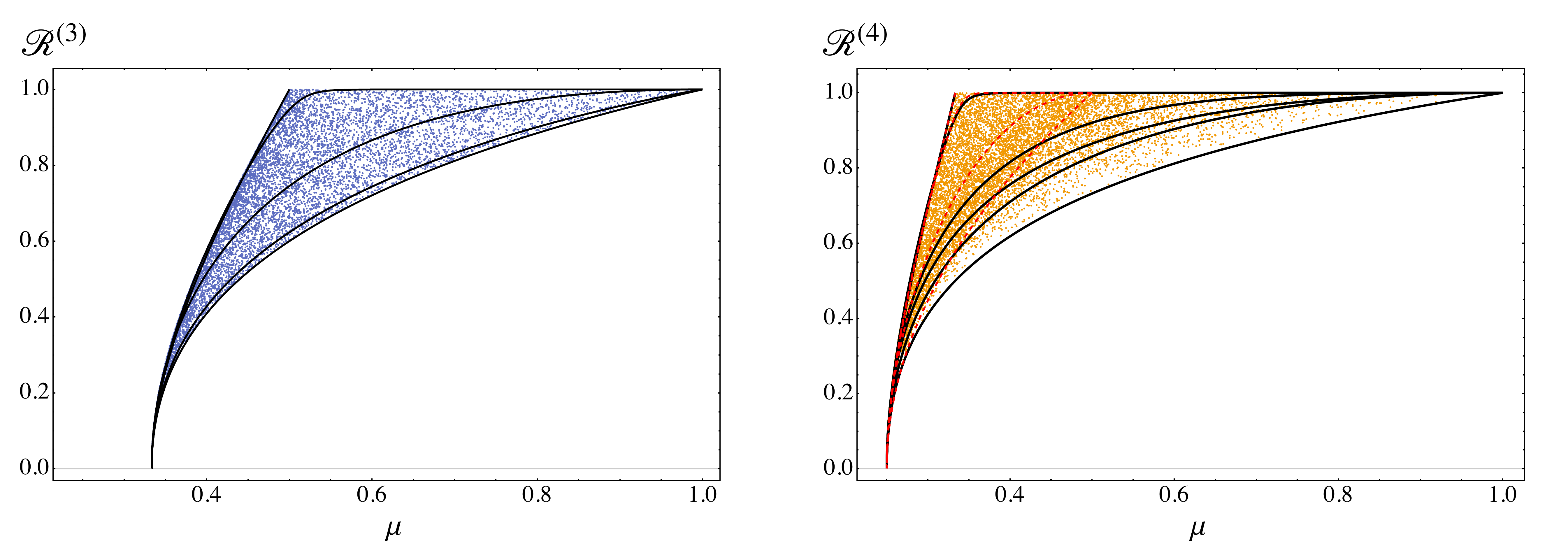}
\caption{Left panel: scatter plot of the AI measure (y-axis) vs purity (x-axis) for qutrit systems. The black lines are the AI for fixed spectrum of $H$, see main text for details. The most left line refers to a choice of $H$ with $2$ degenerate eigenvalues. \\
Right panel: scatter plot of the AI measure (y-axis) vs purity (x-axis) for $4$-dimensional quantum systems. Again, the black lines are the AI for a fixed spectrum of $H$. The red-dashed lines refers to choice of $H$ with at least $2$ degenerate eigenvalues.}
\label{fig:AIrandom}
\end{figure}

Let us now move back to consider finite-dimensional quantum systems described via Hilbert spaces with dimension $d$ larger than two. The most natural way to describe the most general quantum states in this scenario is by the following parametrization 
\begin{equation}
	\varrho_{\bm\gamma} = \frac{1}{d} \left( \mathbbm{1} + \sum_{j=1}^{d^{2}-1}\gamma_{i}\Sigma_{i}\right) \,
\end{equation}
where the matrices $\Sigma_i$ denotes the {generators of the Lie algebra $\mathfrak{su}(d)$}, and thus correspond to Pauli matrices for $d=2$ and to Gell-Mann matrices for $d=3$. The $\mathfrak{p} =d^{2}-1$ coordinates $\bm\gamma$ represent the normalized Cartesian vector in the $d^{2}-1$-dimensional Bloch space and are also known as the components of the $d^{2}-1$-dimensional Bloch vector or as mixture coordinates \cite{bengtsson2017geometry}. In the following we consider the estimation of the parameters $\bm\gamma$, i.e. the full tomography of the state. It is known that this model is a $\mathcal{D}$-invariant model and hence the Holevo bound is equal to the right logarithmic derivative scalar bound \cite{huangjun2012quantum}.\\
Analytical solutions for the AI measure are very hard to derive with arbitrary dimension $d$. Hence, we address the problem numerically by randomly generating quantum density matrices corresponding to $d=3$, $d=4$ and $d=5$. The method we used to generate random density matrix follows two steps: 
\begin{enumerate}
\item First, we generate the eigenvalues of the density matrix $\varrho_{\bm\gamma}$, which belongs to the probability simplex $\bm{X}_{d} = \{\bm{x} =(x_{1},...,x_{d})\vert \sum_{i=1}^{d} x_{i} = 1\}$.
\item Second, we randomly generate unitary matrix $U$. We remind that different $U$ corresponds to different eigenvectors of $\varrho_{\bm\gamma}$ and thus to different quantum states and more importantly to different values of the parameters $\bm{\gamma}$.
\end{enumerate}
For each random state we evaluate the AI measure from the definition \eqref{eq:Rdefinition} by numerically solving the Lyapunov equation for the SLD operators and then by evaluating the SLD-QFI and Uhlmann curvature matrices. The results are reported in Fig. \ref{fig:AIrandom} as function of the purity of the state. We clearly see that there is no one-to-one correspondence between the value of $\mu$ and the value of $\mathcal{R}^{(d)}$. However, we see that the AI is strictly less than one whenever the purity of the state is lower than $\mu = 1/(d-1)$. Thus, we can state the following conjecture:\\

\noindent
{\bf Conjecture 1}. {\em Given a quantum statistical model described by a family quantum states $\varrho_{\lambdaB}$ living in a $d$-dimensional Hilbert space, the maximum amount of the AI measure $\mathcal{R}_{\lambdaB} = 1$ may be observed only if the purity of the quantum state is larger than a minimum threshold $\mu_{\sf min} = 1 / (d -1)$.} \\

\noindent
This conjecture is clearly consistent for the two cases discussed above, i.e. qubit and single-mode Gaussian states, whose AI was exactly derived, and it is surrogated by the numerical evidence in Fig. \ref{fig:AIrandom}.
\par
We now introduce a second way to parametrize a $d$-dimensional quantum system that will be particularly useful fo our purposes, i.e.
\begin{equation}
\label{eq:expcoord}
	\varrho_{\bm\lambda} = \frac{e^{-\beta H}}{\Tr\left[e^{-\beta H}\right]}, \quad \quad H = \sum_{j=1}^{d^{2}-1} \lambda_{j} \Sigma_{j}
\end{equation}
The coordinates $\bm\lambda$ are known as exponential coordinates \cite{bengtsson2017geometry} while the parameter $\beta$ play the role of the inverse temperature of the state with respect to the Hamiltonian operator $H$, clearly recalling the Gibbs state form used in statistical mechanics. However, this is not a \emph{true} coordinate since it can be included in the definition of the $\bm{\lambda}$, hence must be considered more as a rescaling parameter which allow us to constrain the exponential coordinates to be in a finite interval ($| \lambda_{i} |\leq 1$ in our simulations).
\par
In general, a closed formula for the AI is hard to derive also with this parameterization. Hence, we investigate the problem by considering a simplified case, i.e. a diagonal $H = \textup{diag}\left\{\Delta_{0},\Delta_{1},...,\Delta_{d-1}\right\}$, with the constraint $| \Delta_i | \leq 1$. In other words, we set $\lambda_{j}=0$ for the indexes $j$ corresponding to non-diagonal operators $\Sigma_{j}$, and we address the estimation for all the parameters corresponding to the exponential coordinates $\bm\lambda$. The reason why we study this class of states is twofold: the first is that the calculations are greatly simplified, the second is that the results we find seem to be valid more generally, as we will explain later in more detail.\\
We have evaluated analytically the AI measure corresponding to the estimation of all the $d^2-1$ parameters encoded in this particular quantum statistical model, finding
\begin{equation}
\label{eq:AIquditconj}
	\mathcal{R}_{\sf qudit} = \tanh\left(\frac{\beta\Delta_{M}}{2}\right)
\end{equation}
where $\Delta_{M} = \max_{i}{\Delta_{i}}-\min_{j}{\Delta_{j}}$, i.e. the maximum difference between the eigenvalues of $H$ (we remark that the calculations are much simpler by evaluating the matrices for the parameters $\bm\gamma$, and exploiting the fact that $\mathcal{R}_{\bm\lambda}$ is invariant under reparametrization). The result in \eqref{eq:AIquditconj} is consistent with the AI for a qubit states, as for $d=2$ the purity $\mu$ and the argument of the $\tanh$ are in one-to-one correspondence. Indeed, the purity can be written in terms of $\bm\lambda$ as $\mu = (1+\tanh(\beta \vert \bm\lambda \cdot\bm\lambda\vert)^{2})/2$. In the case of a diagonal $H$, we have $\lambda_{1}=\lambda_{2}=0$ and $\Delta_{M} = 2\vert\lambda_{3}\vert$, and thus 
\begin{equation}
	\mu = \frac{1}{2}(1+\tanh(\beta \vert \lambda_{3}\vert)^{2} = \frac{1}{2}\left(1+\tanh\left( \frac{\beta\Delta_{M}}{2}\right)^{2}\right) \,.
\end{equation}
By simply using \eqref{eq:AIquditconj} we indeed obtain the result in Eq.~\eqref{eq:AIqubit}. Remarkably, also the Gaussian case might be reduced to this expression if we consider that the purity in terms of the average number of excitation $N$ is $\mu=(2N+1)^{-1}$. By replacing the value of $N$ by the formula for the Bose-Einstein statistics $N= 1/(\exp{\beta \omega}-1)$, we obtain that the purity can be written as $\mu = \tanh(\beta\omega/2)$. Then, the AI \eqref{eq:AIGauss} becomes exactly equal to \eqref{eq:AIquditconj} by fixing $\Delta_{M} = 2\omega$.
\par 
This line of reasoning of course does not prove that the AI is equal to \eqref{eq:AIquditconj} for arbitrary $d$-dimensional density matrix $\varrho_{\bm\lambda}$. To try to answer this question, we addressed the problem by studying these quantities for the numerically generated random quantum states of dimension $d=3,4,5$. In more detaiil
\begin{enumerate}
\item We evaluate the quantity $\beta \Delta_{M}$  for each random states $\varrho_{\bm\lambda}$ we generate (see previous paragraph for the details of the random generation of $\varrho_{\bm\lambda}$) via the formulas $\beta^{(rand)} \Delta^{(rand)}_{i} = -\log(x_{i})$ , and $\beta^{(rand)}\Delta_{M}^{(rand)} =\max_{i}\beta^{(rand)}\Delta^{(rand)}_{i}-\min_{j}\beta^{(rand)}\Delta^{(rand)}_{j}$ (we remind that the quantities $\{x_i\}$ correspond to the eigenvalues of $\varrho_{\lambdaB}$).
\item For each random state we evaluate the AI measure corresponding to its full estimation via Eq. \eqref{eq:Rdefinition}, and we compare it with the formula \eqref{eq:AIquditconj} evaluated via the parameter $\beta^{(rand)}\Delta_{M}^{(rand)}$.
\item We find that the two quantities match for all the quantum states that we have numerical generated.
\end{enumerate} 
We conclude that for $d=3,4,5$ the AI is fully determined only by $\beta \Delta_{M}$, i.e. it corresponds to a property of the spectrum of $H$ only (to be more precise it depends only on the maximum and minimum eigenvalues of $H$).
For these reasons, we state a second conjecture here \\

\noindent
{\bf Conjecture 2}. {\em Given a quantum statistical model described by a family quantum states $\varrho_{\lambdaB}$ living in a $d$-dimensional Hilbert space, the AI corresponding to the full estimation of the state is given by \eqref{eq:AIquditconj}, where  $\beta$ is the fictious temperature of the state with respect to the Hamiltonian operator $H$ in the exponential coordinates \eqref{eq:expcoord}.} \\

\noindent
A further element in favour of this conjecture comes from Fig. \ref{fig:AIrandom}, in which we clearly see that the AI for a fixed spectrum (black lines) span the whole area covered by the AI measure values corresponding to the random states generated. In addition, we see that the AI for a $4$-dimensional system with $H$ having two eigenvalues with degeneracy $2$ reaches the maximum value of $\mathcal{R}=1$ when the purity is $\mu =1/2$, since in the case in which $\Delta_{0}=\Delta_{1}$ and $\Delta_{2}=\Delta_{3}$, we have that $\mu \to 1/2$ for $\beta \to \infty$. \\
We finally remark again that, as we described above, the formula in Eq. (\ref{eq:AIquditconj}) can be readily evaluated, after diagonalizing the quantum state describing the quantum statistical model $\varrho_{\lambdaB} = \sum_i x_i |\psi_i\rangle\langle \psi_i |$, and by observing that $\beta \Delta_i = -\log(x_i)$.

\section{Further properties of asymptotic incompatibility measure} \label{s:properties}
In this section we analyze in more detail some further properties of the AI measure $\mathcal{R}_{\lambdaB}$, sheding more light on the relationship between a given quantum statistical model $\varrho_{\lambdaB}$ and possible {\em sub-models} that arise when one or more parameters of the set $\lambdaB$ are considered known and thus not to be estimated.\\
In the definition of $\mathcal{R}_{\lambdaB}$ in Eq. (\ref{eq:Rdefinition}), only the the largest eigenvalue of the matrix $\bm{\mathcal{I}}= i \,\bm{\mathcal{Q}}(\bm{\lambda})^{-1} \bm{\mathcal{U}}(\boldsymbol{\lambda})$ is considered. In the following we will study more in detail the role of the spectrum of $\bm{\mathcal{I}}$ in the characterization of the quantum statistical model. To do so, we first introduce the Cauchy interlace theorem \cite{hwang2004cauchy}:
\begin{theorem}[Cauchy interlace theorem]
Consider a $N\times N$ hermitian matrix $\bm{\mathcal{A}}$ and any $(N-1)\times(N-1)$ principal sub-matrix $\bm{\mathcal{B}}$ of $\bm{\mathcal{A}}$. Consider the corresponding eigenvalues in decreasing order $\bm{a} = \{a_1,\ldots,a_N\}$ and  $\bm{b} = \{b_1,\ldots,b_{N-1}\}$. Then, $\bm{a}$ interlace $\bm{b}$, that is 
\begin{eqnarray}
	a_N \leq b_{N-1} \leq a_{N-1} \leq \ldots \leq b_2 \leq a_2 \leq b_1 \leq a_1. 
\end{eqnarray}
\label{th:cauchy}
\end{theorem}

\noindent
This result can now be applied for our purposes to the matrix $\bm{\mathcal{I}}$, leading to the following theorem:
\begin{theorem}[Bound on AI measure for quantum statistical sub-models]
Given a quantum statistical model defined by a set of $\mathfrak{p}-$parameters $\lambdaB$, and any other possible sub-model defined by a set of (possibly reparametrized) $(\mathfrak{p}-1)$ parameters $\tilde{\bm{\lambda}}$, then the two corresponding AI measures $\mathcal{R}^{(\mathfrak{p})}_{\bm{\lambda}}$ and $\mathcal{R}^{(\mathfrak{p-1})}_{\tilde{\bm{\lambda}}}$ satisfy the inequality
\begin{align}
 	r_{2} \leq \mathcal{R}^{(\mathfrak{p}-1)}_{\tilde{\bm{\lambda}}} \leq \mathcal{R}^{(\mathfrak{p})}_{\bm\lambda} \,,
\end{align}
where $r_{2}$ denotes the second largest eigenvalue of the matrix $\bm{\mathcal{I}}$ corresponding to the quantum statistical model $\varrho_{\lambdaB}$. \\
{\bf Proof}. As observed in \cite{carollo2019quantumness}, the eigenvalues of $\bm{\mathcal{I}}$ are either $0$ or given in pairs $h_i = \pm r_i$, with $0<r_{i}\leq 1, i=1,\ldots,\delta$ and $\delta \leq \left \lfloor{(\mathfrak{p}+1)/2}\right \rfloor $. The thesis of the theorem is thus a simple corollary of the Cauchy interlace theorem \ref{th:cauchy} stated above. \\
\end{theorem}

One can further show that, if we consider smaller sub-matrices of $\bm{\mathcal{I}}$, we can recursively apply this argument to smaller statistical sub-model $\bar{\bm\lambda}$ with $(\mathfrak{p}-j)$ parameters. Eventually one obtains that the AI measure for the corresponding statistical sub-model is bounded as
\begin{eqnarray}	
	r_{j+1} \leq \mathcal{R}^{(\mathfrak{p}-j)}_{\bar{\bm{\lambda}}} \leq \mathcal{R}^{(\mathfrak{p})}_{\bm\lambda}.
\end{eqnarray}	
In particular we see that any statistical sub-model with $(\mathfrak{p}-j)$ parameters is incompatible if we restrict ourselves to $j\leq \delta-1$, since for those $j$ we have that $r_{i} >0$. This observation leads to the following corollary: \\

\noindent
{\bf Corollary} (Upper bound on the number of compatible parameters of a quantum statistical model). {\em By denoting with $\delta$ the number of strictly positive eigenvalues of $\bm{\mathcal{I}}$, then the quantum statistical model $\varrho_{\lambdaB}$ contains at number of compatible parameters that is upper bounded as 
\begin{align}
\mathfrak{p}_{\sf comp} \leq \mathfrak{p}^{(bound)}_{\sf comp} = \mathfrak{p}-\delta \,.
\end{align}}\\

\noindent
This result can be easily applied to the full estimation problem we studied in the previous section. Indeed, we can use the same evidence we have for the Conjecture 2 to conjecture that the eigenvalues of the matrix $\bm{\mathcal{I}}$ are
\begin{equation}
	\text{Eig}(\bm{\mathcal{I}}) = \left\{0,...,0, \pm r_{(1,0)}, \pm r_{(2,0)}, \pm r_{(2,1)} ,..., \pm r_{(d-1,d-3)}, \pm r_{(d-1,d-2)} \right\}
\end{equation}
where
\begin{equation}
	r_{i,j} = \tanh\left(\frac{\beta(\Delta_{i}-\Delta_{j})}{2}\right)
\end{equation}
and $(\Delta_{0},...,\Delta_{d-1})$ are the $\beta$-normalized eigenvalues in ascending order of $H$, defined in Eq. \eqref{eq:expcoord}. If the eigenvalues of $H$ are non-degenerate, the number of strictly positive eigenvalues of $\bm{\mathcal{I}}$ is simply
\begin{equation}
	\delta_{\sf ndg} = \begin{pmatrix} d \\ 2 \end{pmatrix}=\frac{d(d-1)}{2}.
\end{equation} 
Instead, for $H$ with degenerate spectrum, the result slightly change. Indeed, if we denote with $\kappa$ the number of distinct degenerate eigenvalues and with $\eta_{i}$ the corresponding degeneracies (with $i = 1,...,\kappa$), then the number of strictly positive eigenvalues is 
\begin{equation}
	\delta_{\sf dg} = \begin{pmatrix} d \\ 2 \end{pmatrix} - \sum_{i=1}^{\kappa} \begin{pmatrix} \eta_{i} \\ 2 \end{pmatrix} = \frac{d(d-1)}{2}-\sum_{i=1}^{\kappa}\frac{\eta_{i}(\eta_{i}-1)}{2}.
\end{equation}
Thus, given the numerical evidence we obtained from our simulations, we conjecture the following\\

\noindent
{\bf Conjecture 3} (Upper bound on the number of compatible parameters for the full estimation of a $d$-dimensional quantum system). {\em By denoting with $d$ the dimension of the quantum system $\varrho_{\bm\lambda}$, the number of compatible parameters $\mathfrak{p}_{\sf comp}$ in the full estimation of the $d^2-1$-parameters describing $\varrho_{\bm\lambda}$ is upper bounded by the quantity
\begin{align}
\label{eq:maxcompfullqudit}
	\mathfrak{p}^{(bound)}_{\sf comp} & = d^{2}-1 -  \delta_{\sf dg}  \nonumber\\
	& = \frac{(d+2)(d-1)}{2} + \sum_{i=1}^{\kappa}\frac{\eta_{i}(\eta_{i}-1)}{2},
\end{align}
where $\kappa$ is the number of distinct degenerate eigenvalues and $\eta_{i}$ the degeneracy degree, with $i=1,...,\kappa$.}\\

\noindent
This conjecture shows that the value of $\mathfrak{p}^{(bound)}_{\sf comp}$ depends on the values of the parameters, and in particular on the values that makes the corresponding density matrix degenerate. Indeed, the larger is the number of degenerate eigenvalues, the larger is the maximum number of possible compatible parameters. In the limit of full degeneracy, i.e. $\kappa = 1$ and $\eta_{i}=d$, we see that $\mathfrak{p}^{(bound)}_{\sf comp} \to d^{2}-1$, i.e equals the number the number of parameters of the model. Instead, in the case with no degeneracy in the spectrum of $H$, we have the minimum value, i.e. $\mathfrak{p}^{(bound)}_{\sf comp} = (d+2)(d-1)/2$. We would also like to stress that the model is always full-rank as far as the value of $\beta$ is finite, independently on the number of degenerate eigenvalues of $H$.
\par
An upper bound on the number of compatible parameters has been also derived recently in \cite{kukita2020} by following a different approach and exploiting the algebraic structure of the quantum statistical model. We have compared this bound with our bound for the full estimation of a qubit, that is by considering the estimation of the Bloch sphere parameters $\bm\lambda=(r,\theta,\phi)$. We found that the matrix ${\bm{\mathcal{I}}}$ has eigenvalues ${\rm Eig}(\bm{\mathcal{I}}) = \{r , -r ,0 \}$, leading to the upper bound $\mathfrak{p}^{(bound)}_{\sf comp} = 2$. This results does indeed coincide with the one obtained in \cite{kukita2020}.
\par
\begin{figure}
	\centering
	\includegraphics[width=1\textwidth]{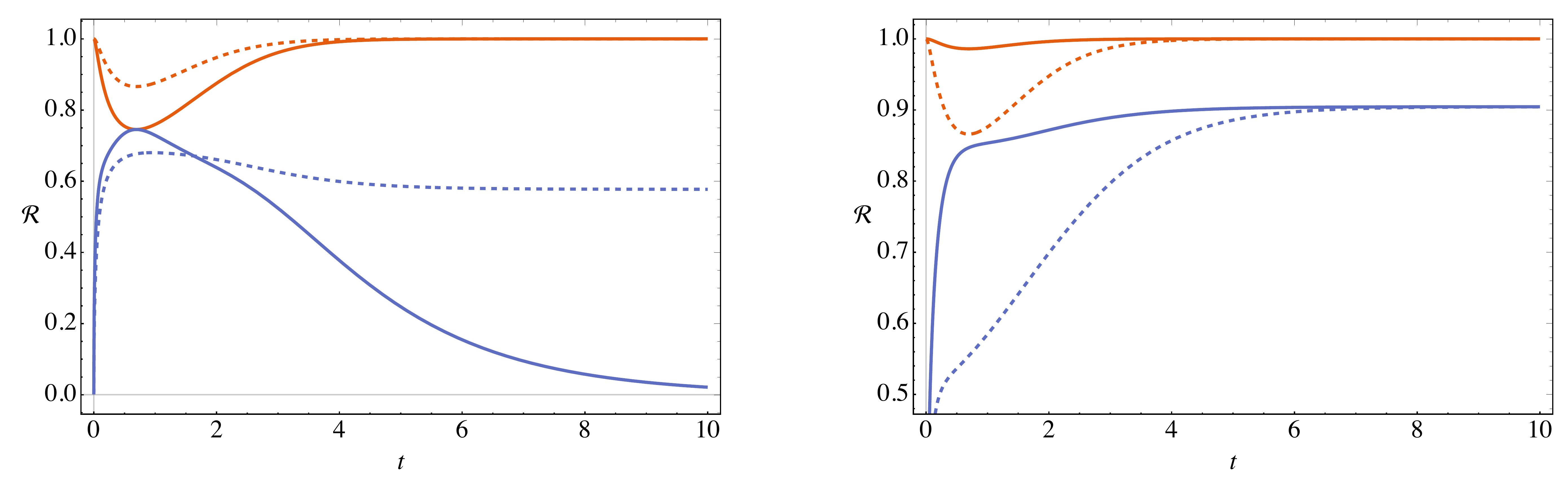}
	\caption{Plot of $\mathcal{R}^{(5)}_{\bm{\lambda}}$ (red line) and $\mathcal{R}^{(2)}_{\ddot{\bm{\lambda}}}$ (blue line). Left panel: $\langle \hat{a}^{\dagger}\hat{a}\rangle = 4$; thick line: $\eta=1$; dashed line: $\eta=0.5$. Right panel: $\eta=0.1$; thick line $\langle \hat{a}^{\dagger}\hat{a}\rangle = 4$; dashed line: $\langle \hat{a}^{\dagger}\hat{a}\rangle = 20$. Both panel: $\gamma=\omega=1$.}
	\label{fig:comparisonR}
\end{figure}

We also remark that, while the bound in \cite{kukita2020} was derived for finite dimensional Hilbert space, in our case we made no assumption on the dimension of the Hilbert space. Hence we can apply our result also for infinite dimensional system. For instance, let us consider the paradigmatic case of the estimation of the parameters characterizing a single-mode Gaussian state defined as in Eq. (\ref{eq:GState}) and already treated in Sec. \ref{s:AIandPurity}. The spectrum of the corresponding matrix $\bm{\mathcal{I}}$ is given by
\begin{eqnarray}
	\textup{Eig}(\bm{\mathcal{I}}) = \left\{\frac{2\mu}{1+\mu^2},\mu,0,-\mu,-\frac{2\mu}{1+\mu^2}\right\}.
\end{eqnarray}
This leads to the AI measure for the complete statistical model $\mathcal{R}^{(5)}_{\bm\lambda}=2\mu/(1+\mu^2)$ we discussed above. By further inspecting the set of eigenvalues above we can however also conclude that for any subset of $\mathfrak{p}=4$ parameters $\tilde{\bm{\lambda}}$, the corresponding AI parameter is bounded as
\begin{eqnarray}
	\mu \leq \mathcal{R}^{(4)}_{\tilde{\bm\lambda}} \leq \frac{2\mu}{1+\mu^2},
\end{eqnarray}
and that there is a maximum of $\mathfrak{p}_{\sf comp}^{(bound)}=3$ compatible parameters. In general, by observing the form of the Uhlmann curvature matrix we also see that the only incompatible models are the one which deals with the simultaneous estimation of $\{r,\varphi\}$ or $\{\mathfrak{Re}{\alpha},\mathfrak{Im}{\alpha}\}$. \\
In the examples above, if we restrict to subsets of the original parameters $\lambdaB=\{\mathfrak{Re}{\alpha},\mathfrak{Im}{\alpha},r,\varphi,N\}$, all the results are directly availabe from the explicit solution of the SLD-QFIM \eref{eq:Qmatelements1}~--~\eref{eq:Qmatelements6} and the Uhlmann matrix \eref{eq:Umatelements1}~--~\eref{eq:Umatelements2}. Here below we rather present an example where the results cannot be obtained straightforwardly in analytical form, while the properties discussed above can still provide a first insight without needing to solve exactly the estimation problem.\\
We address the estimation of two dynamical parameters, the frequency and the loss rate $\ddot{\bm{\lambda}} = \{\omega,\gamma\}$ that characterize the evolution due to the Markovian master equation
\begin{eqnarray}
	\dot{\varrho} = -i\frac{\omega}{2}[\hat{q}^2 + \hat{p}^2,\varrho ]  + \gamma \mathcal{D}[\hat{a}]\varrho,	
\end{eqnarray} 
with $\mathcal{D}[\hat{a}]\varrho = \hat{a}\varrho \hat{a}^\dag - \frac{1}{2}\left(\hat{a}^\dag\hat{a}\varrho+\varrho\hat{a}^\dag \hat{a} \right)$. 
By consider an initial Gaussian state $\varrho(0)$ as the one in Eq. (\ref{eq:GState}), the dynamics remains Gaussian and thus can be fully described by means of the first and second moments of its quadrature operators. In particular one can readily evaluate analytically the purity of the quantum state, obtaining
\begin{eqnarray}
	\mu(t) = \sqrt{(1-e^{-\gamma t})(1-e^{-\gamma t}(1-2 s_2(0)))+e^{-2\gamma t}\mu(0)^2},
\end{eqnarray}
>From this result and via Eq. (\ref{eq:AIGauss}), we can directly evaluate $\mathcal{R}^{(5)}_{\bm{\lambda}}$, whereas the AI measure $\mathcal{R}^{(2)}_{\ddot{\bm{\lambda}}}$ for the quantum statistical model defined above is evaluated numerically by exploiting the techniques reported in \ref{a:gaussian}. To present the results, we choose the initial state to be a pure state $N=0$ and we parametrize it in terms of its average excitations number $\langle \hat{a}^\dag \hat{a} \rangle$ and its fraction of squeezing $\eta$ defined as
\begin{eqnarray}
	\langle \hat{a}^\dag \hat{a} \rangle = \vert \alpha \vert^2 + \sinh(r)^2, \\
	\eta = \frac{\sinh(r)^2}{\langle\hat{a}^\dag \hat{a}\rangle} \,.
\end{eqnarray} 
We plot the results in \Fref{fig:comparisonR} and we see that $\mathcal{R}^{(5)}_{\bm{\lambda}}$ is, as expected, always larger than $\mathcal{R}^{(2)}_{\ddot{\bm{\lambda}}}$, confirming that $\mathcal{R}^{(5)}_{\bm{\lambda}}$ can be used as an upper bound for the quantumness for any other sub-statistical model. Besides, as we see from the left panel in \Fref{fig:comparisonR}, the upper bound may not be much informative, since the evolution of $\mathcal{R}^{(5)}_{\bm{\lambda}}$ and $\mathcal{R}^{(2)}_{\ddot{\bm{\lambda}}}$ in time are opposite. In addition, this example shows that the condition on the number of eigenvalue is not sufficient to have a compatible model.

\section{Conclusions}\label{s:conclusion}
In this paper, we have studied in details the properties of the AI measure $\mathcal{R}_{\bm{\lambda}}$ of incompatibility and discussed its use in assessing the quantumness of  multiparameter estimation problems. At first, we have focused on the estimation of the full set of parameters characterizing a given quantum system, showing that for qubits
and  single-mode Gaussian systems $\mathcal{R}_{\bm{\lambda}}$ is a simple monotonous function of the purity $\mu$ of the state. We have then considered a generic $d$-dimensional quantum systems and, using analytical and numerical tools, we found that in general the one-to-one correspondence between $\mathcal{R}_{\bm{\lambda}}$ and $\mu$ does not hold anymore. However, numerical results suggest that the maximum number of AI is attainable only for quantum statistical models having a purity $\mu \geq 1/(d-1)$. We conjecture that this may be a general property of $d$-dimensional quantum systems. Upon considering quantum statistical models described as Gibbs state of a given Hamiltonian, we have also shown that the AI measure can be written as a simple function of the {\em fictitious} temperature parameter and the spectrum of the Hamiltonian. 
\par
We have then studied in detail the role of the spectrum of the matrix 
$\bm{\mathcal{I}} = i\bm{\mathcal{Q}}(\bm{\lambda})^{-1}\bm{\mathcal{U}}(\bm{\lambda})$ and have determined bounds on $\mathcal{R}_{\bm{\lambda}}$ for quantum statistical {\em submodels}. In particular, we have shown that the number of strictly positive 
eigenvalues of $\bm{\mathcal{I}}$ determines the maximum number of compatible parameters in a given statistical model. 
\par
Very recently there has been much interest in deriving bounds that apply when one allows measurements on a finite number of copies \cite{Conlon2021,chen2021}; we thus expect that our approach can be extended from the asymptotic model, identifying and studying a hierarchy of incompatibility measures in this {\em finite copies} scenario.
More in general our results pave the way to a deeper understanding of the fundamental properties of multiparameter quantum estimation, and provide potentially useful tools to approach multiparameter problems that cannot be addressed analytically. 
\section*{Acknowledgements}
The authors acknowledge several useful discussions with Francesco Albarelli.
MGAP is member of INdAM-GNFM.
\appendix

\setcounter{section}{0}

\section{The Holevo Cram\'er-Rao bound}
\label{app:HCRB}
The Holevo Bound $C^H(\bm\lambda,\bm{\mathcal{W}})$ can be evluated via the following minimization
\begin{eqnarray}
   	C^H(\bm\lambda,\bm{\mathcal{W}}) = \min_{\hat{\bm{\mathcal{X}}}\in \mathbb{X}_{\bm{\lambda}}}\left[\Tr\left[\bm{\mathcal{W}}\mathfrak{Re}{\bm{\mathcal{Z}}[\hat{\bm{\mathcal{X}}}]}\right] + \Vert \sqrt{\bm{\mathcal{W}}}\mathfrak{Im}{\bm{\mathcal{Z}}[\hat{\bm{\mathcal{X}}}]}\sqrt{\bm{\mathcal{W}}}\Vert_1 \right], 
\end{eqnarray}   
where the elements of the hermitian $d\times d$ matrix $\bm{\mathcal{Z}}$ are
\begin{eqnarray}
	\mathcal{Z}[\hat{\bm{\mathcal{X}}}]_{\mu\nu} = \Tr[\varrho_{\bm{\lambda}} \hat{\mathcal{X}}_\mu\hat{\mathcal{X}}_\nu],
\end{eqnarray}
while $\hat{\bm{\mathcal{X}}}$ is an array of $d$ Hermitian operators such that
\begin{eqnarray}
	\Tr[\varrho_{\bm{\lambda}} \hat{\mathcal{X}}_\mu] = 0\\
	\Tr[\hat{\mathcal{X}}_\mu\partial_\nu\varrho_{\bm{\lambda}}] = \frac{1}{2}\Tr[\varrho_{\bm{\lambda}}\{\hat{\mathcal{X}}_\mu,\hat{L}^S_\nu\}] = \delta_{\mu\nu}
\end{eqnarray}

\section{Multiparameter estimation for continuous-variable Gaussian states}
\label{a:gaussian}
In this section we focus on estimation problems in bosonic continuous variable systems. In particular, we study $n$ modes Gaussian state \cite{ferraro2005gaussian,serafini2017quantum}, which are properly described in terms of quadrature operators $\hat{\bm{r}} = \{\hat{q}_1,\hat{p}_1,\ldots,\hat{q}_n,\hat{p}_n\}$. They satisfy the canonical commutation relation
\begin{eqnarray}
	\left[\hat{r}_j,\hat{r}_k\right] = i \Omega_{jk},
\end{eqnarray}
where $\Omega = i \sigma_y^{\oplus n}$.  A Gaussian state $\varrho$ is completely determined by its vector of first moments $\bm{d} = \Tr[\varrho \bm{\hat{r}}]$ and its covariance matrix (CM) $\bm\sigma = \Tr[\varrho\{(\bm{\hat{r}}-\bm{d}),(\bm{\hat{r}}-\bm{d})^T\}]$, with $\{\hat{\bm{l}},\hat{\bm{l}}^T\}_{jk} = \hat{l}_j\hat{l}_k+\hat{l}_k\hat{l}_j$ \cite{ferraro2005gaussian,serafini2017quantum}.
\par
In order to derive the SLD-QFI matrix elements, several approaches have been proposed in the literature \cite{Pinel2013, Zhang2014,Banchi2015, serafini2017quantum, nichols2018multiparameter, Safranek2018} . As we are interested also in the Uhlmann curvature matrix, in the following we will focus on the derivations pursued in \cite{serafini2017quantum,nichols2018multiparameter} that are indeed based on the writing the SLD operators in terms of the moments of the Guassian states defining the quantum statistical model.
Let us then assume that a set of parameters $\bm\lambda$ defines a quantum statistical models in a family of Gaussian quantum states $\varrho_{\bm\lambda}$. One shows that the SLD for a parameter $\alpha\in\bm\lambda$ is at most quadratic in the moments \cite{serafini2017quantum} and can be written as
\begin{eqnarray}
	\hat{L}^S_\alpha = L^{(0)}_\alpha \mathbb{I} + \bm{L}_\alpha^{(1)}\cdot\hat{\bm{r}} + \hat{\bm{r}}^T\cdot \bm{L}^{(2)}_\alpha\cdot \bm{\hat{r}},
\end{eqnarray}
where $L^{(0)}_\alpha$ is a real number, $\bm{L}^{(1)}_\alpha$ is $2n$ real vector and $\bm{L}^{(2)}_{\alpha}$ is a $2n\times2n$ real symmetric  matrix. After some algebra, one can find that the SLD-QFI matrix elements are given by \cite{serafini2017quantum,nichols2018multiparameter}
\begin{eqnarray}
	\label{eq:QFIMgaussian}
	\mathcal{Q}_{\alpha \beta} = \frac{1}{2} \textup{Tr}[L^{(2)}_\alpha(\partial_\beta \bm\sigma)] + 2 (\partial_\alpha \bm{d}^T) \bm\sigma^{-1} (\partial_\beta \bm{d})
\end{eqnarray}	
Here below we derive the analogous result for the Uhlmann curvature matrix. First, we evaluate the commutator of $\hat{L}^S_\alpha$ and $\hat{L}^S_\beta$, that is
\begin{eqnarray}
\fl
	\left[\hat{L}^S_\alpha,\hat{L}^S_\beta\right] = \left[\bm{L}^{(1)^T}_\alpha \hat{\bm{r}},\bm{L}^{(1)^T}_\beta \hat{\bm{r}}\right] + \left[\bm{L}^{(1)^T}_\alpha \hat{\bm{r}} ,\hat{\bm{r}}^T \bm{L}^{(2)}_\beta \hat{\bm{r}}\right] + \left[\hat{\bm{r}}^T \bm{L}^{(2)}_\alpha \hat{\bm{r}},\bm{L}^{(1)^T}_\beta \hat{\bm{r}}\right] + \nonumber\\
	+  \left[\hat{\bm{r}}^T \bm{L}^{(2)}_\alpha \hat{\bm{r}},\hat{\bm{r}}^T \bm{L}^{(2)}_\beta \hat{\bm{r}}\right].
\end{eqnarray}
By recalling that $\left[\hat{r}_j,\hat{r}_k\right]= i\Omega_{jk}$, the first term of the latter is (we use Einstein convention on indexes summations)
\begin{eqnarray}
	\left[\bm{L}^{(1)^T}_\alpha \hat{\bm{r}},\bm{L}^{(1)^T}_\beta \hat{\bm{r}}\right] = L^{(1)}_{\alpha,i}L^{(1)}_{\beta,j}\left[\hat{r}_i,\hat{r}_j\right]=i \bm{L}^{(1)^T}_\alpha \Omega \bm{L}^{(1)}_\beta \mathbb{I};
\end{eqnarray}
the second term is ( we recall that $L^{(2)}_{\alpha,jk} = L^{(2)}_{\alpha,kj}$)
\begin{eqnarray}
	\left[\bm{L}^{(1)^T}_\alpha \hat{\bm{r}} ,\hat{\bm{r}}^T \bm{L}^{(2)}_\beta \hat{\bm{r}}\right] & = L^{(1)}_{\alpha,i}L^{(2)}_{\beta,jk}\left[\hat{r}_i,\hat{r}_j\hat{r}_k\right] = \nonumber \\
	& = L^{(1)}_{\alpha,i}L^{(2)}_{\beta,jk}\left(\left[\hat{r}_i,\hat{r}_j\right]\hat{r}_k+\hat{r}_j\left[\hat{r}_i,\hat{r}_k\right]\right) =\nonumber \\
	& = L^{(1)}_{\alpha,i}L^{(2)}_{\beta,jk}\left(i\Omega_{ij}\hat{r}_k+\hat{r}_ji\Omega_{ik}\right) = \nonumber\\
	& = 2i \bm{L}^{(1)^T}_\alpha \Omega \bm{L}^{(2)}_{\beta} \hat{\bm{r}};
\end{eqnarray}
while the third term is 
\begin{eqnarray}
	\left[\hat{\bm{r}}^T \bm{L}^{(2)}_\alpha \hat{\bm{r}},\bm{L}^{(1)^T}_\beta \hat{\bm{r}}\right] = 2 i \hat{\bm{r}}^T\bm{L}^{(2)}_\alpha \Omega \bm{L}^{(1)}_\beta;
\end{eqnarray}
and eventually the fourth term is (we used the symmetry of $L^{(2)}_{\alpha}$ and the skew-symmetry of $\Omega$)
\begin{eqnarray}
\fl
	\left[\hat{\bm{r}}^T \bm{L}^{(2)}_\alpha \hat{\bm{r}},\hat{\bm{r}}^T \bm{L}^{(2)}_\beta \hat{\bm{r}}\right] & = L^{(2)}_{\alpha,ij}L^{(2)}_{\beta,kl}\left[\hat{r}_i\hat{r}_j,\hat{r}_k\hat{r}_l\right] = \nonumber \\
	& = L^{(2)}_{\alpha,ij}L^{(2)}_{\beta,kl} \left(i \Omega_{jk} \hat{r}_i \hat{r}_l + i \Omega_{ik} \hat{r}_j \hat{r}_l +i \Omega_{jl}\hat{r}_k\hat{r}_i + i\Omega_{il} \hat{r}_k \hat{r}_j\right) = \nonumber \\
	& = 2i \bm{\hat{r}}^T \left(\bm{L}^{(2)}_\alpha \Omega \bm{L}^{(2)}_\beta - \bm{L}^{(2)}_\beta \Omega \bm{L}^{(2)}_\alpha\right)\bm{\hat{r}}.
\end{eqnarray}
So eventually we obtain
\begin{eqnarray}
	\left[\hat{L}^S_\alpha,\hat{L}^S_\beta\right] = i \left(\bm{L}^{(1)^T}_\alpha \Omega \bm{L}^{(1)}_\beta\mathbb{I} + 2 \bm{B}^T\hat{\bm{r}} + 2\bm{\hat{r}}^T \bm{A} \bm{\hat{r}}\right)
\end{eqnarray}
where $\bm{A}$ and $\bm{B}$ are respectively a $d\times d$ symmetric matrix and a $d$ vector
\begin{eqnarray}
	\bm{A} = \bm{L}^{(2)}_\alpha \Omega \bm{L}^{(2)}_\beta - \bm{L}^{(2)}_\beta \Omega \bm{L}^{(2)}_\alpha = 2  \bm{L}^{(2)}_\alpha \Omega \bm{L}^{(2)}_\beta, \\
	\bm{B} = \bm{L}^{(2)}_\alpha \Omega \bm{L}^{(1)}_\beta - \bm{L}^{(2)}_\beta\Omega \bm{L}^{(1)}_\alpha 
\end{eqnarray}
Now we can evaluate
\begin{eqnarray}
	\mathcal{U}_{\alpha\beta} = -\frac{i}{2} \Tr\left[\varrho_{\bm{\lambda}}\left[\hat{L}^S_\alpha,\hat{L}^S_\beta\right]\right] = \\
	= \frac{1}{2}\bm{L}^{(1)^T}_\alpha \Omega \bm{L}^{(1)}_\beta + \bm{L}^{(1)^T}_\alpha \Omega \bm{L}^{(2)}_{\beta} \bm{d} + \bm{d}^T\bm{L}^{(2)}_\alpha \Omega \bm{L}^{(1)}_\beta + \Tr\left[\varrho_{\bm{\lambda}} \bm{\hat{r}}^T \bm{A}\bm{\hat{r}}\right]
\end{eqnarray}
Since $A_{ij}\Omega_{ij} =0$, we obtain that
\begin{eqnarray}
	\Tr\left[\varrho_{\bm{\lambda}} \bm{\hat{r}}^T \bm{A}\bm{\hat{r}}\right] & = A_{ij} \Tr\left[\varrho_{\bm{\lambda}} \hat{r}_i \hat{r}_j\right] = \nonumber \\
	& = A_{ij}\left(\frac{\sigma_{ij}}{2}+ \frac{i\Omega_{ij}}{2} + d_id_j\right) = \nonumber \\
	& = \frac{1}{2}\Tr\left[\bm{A}\bm\sigma\right] + \bm{d}^T \bm{A} \bm{d} = \nonumber \\
	& = \Tr\left[\bm{L}^{(2)}_\alpha\Omega\bm{L}^{(2)}_\beta\bm\sigma\right] + 2 \bm{d}^T \bm{L}^{(2)}_\alpha \Omega \bm{L}^{(2)}_\beta \bm{d}
\end{eqnarray}
>From \cite{serafini2017quantum} we recall that $\bm{L}^{(1)}_\alpha = 2\bm\sigma^{-1}(\partial_\alpha \bm{d})- 2 \bm{L}^{(2)}_\alpha \bm{d}$ and so we derive the following
\begin{eqnarray}
	\frac{1}{2}\bm{L}^{(1)^T}_\alpha \Omega \bm{L}^{(1)}_\beta &= 2 \bigg((\partial_\alpha \bm{d})^T \bm\sigma^{-1}\Omega\bm\sigma^{-1} (\partial_\beta \bm{d})- \bm{d}^T\bm{L}^{(2)}_\alpha\Omega\bm\sigma^{-1}(\partial_\beta\bm{d}) + \\
	& \quad - (\partial_\alpha \bm{d})^T\bm\sigma^{-1}\Omega L^{(2)}_\beta \bm{d} + \bm{d}^T \bm{L}^{(2)}_\alpha \Omega\bm{L}^{(2)}_\beta\bm{d}\bigg) \\
	\bm{L}^{(1)^T}_\alpha \Omega \bm{L}^{(2)}_{\beta} \bm{d} & = 2 \left((\partial_\alpha \bm{d})^T \bm\sigma^{-1}\Omega \bm{L}^{(2)}_\beta\bm{d}-\bm{d}^T\bm{L}^{(2)}_\alpha\Omega\bm{L}^{(2)}_\beta\bm{d}\right) \\
	\bm{d}^T\bm{L}^{(2)}_\alpha \Omega \bm{L}^{(1)}_\beta & = 2 \left(\bm{d}^T\bm{L}^{(2)}_\alpha\Omega\bm\sigma^{-1}(\partial_\beta \bm{d})- \bm{d}^T\bm{L}^{(2)}_\alpha \Omega \bm{L}^{(2)}_\beta\bm{d}\right)
\end{eqnarray}
By inserting these last calculations, we end up with the following Uhlmann matrix
\begin{eqnarray}
	\mathcal{U}_{\alpha\beta} = \Tr\left[\bm{L}^{(2)}_\alpha\Omega\bm{L}^{(2)}_\beta\bm\sigma\right] + 2 (\partial_\alpha \bm{d}) \bm\sigma^{-1} \Omega \bm\sigma^{-1}(\partial_\beta \bm{d}). \label{eq:Uhlmanngaussian}
\end{eqnarray}
Please notice that our result differ from that in \cite{nichols2018multiparameter} by a factor $2$ in the first term of the Uhlmann matrix. \\

If we now focus on single-mode Gaussian state, we can write \eref{eq:QFIMgaussian} and \eref{eq:Uhlmanngaussian} only in terms of the covariance matrix $\bm\sigma$, the vector $\bm{d}$ and the purity of the state $\mu = 1/\sqrt{\det{\bm\sigma}}$ as
\begin{align}
 	\mathcal{Q}_{\alpha\beta} &= \frac{1}{2} \frac{\textup{Tr}[(\bm{\sigma}^{-1}\partial_\alpha \bm{\sigma}) (\bm{\sigma}^{-1} \partial_\beta \bm{\sigma})]}{1+\mu^2} + \frac{2\partial_\alpha \mu \partial_\beta \mu}{1-\mu^4}  \nonumber \\
	&\,\,\,\,\,\,\,\, + 2\left(\partial_\alpha \bm{d} \right)^T \bm\sigma^{-1} (\partial_\beta \bm{d}) \\
 	\mathcal{U}_{\alpha \beta} &= \frac{\mu^2}{2(\mu^2+1)^2} \textup{Tr}\left\{\bm\sigma\Omega \left[\partial_\alpha \bm{\sigma}\bm{\sigma}^{-1}, \partial_\beta \bm\sigma\bm{\sigma}^{-1}\right]\right\} \nonumber \\
	&\,\,\,\,\,\,\,\,+ 2 \mu^2(\partial_\alpha \bm{d})  \Omega (\partial_\beta \bm{d}). 
\end{align} 

\section*{References}
\bibliographystyle{unsrt}
\bibliography{biblio}

\end{document}